\documentclass[11pt]{article}
\usepackage{amsmath,sint,epsfig,cite,graphics}
\usepackage{macros_static}
\usepackage{textcomp}

\begin{document}

\begin{titlepage}

\begin{flushright}
\vskip 0.7cm
DESY 08-079 \\
SFB/CPP-08-33\\
CERN-PH-TH/2008-138\\
HU-EP-08/21\\
\end{flushright}

\vskip 0.35cm
\begin{center}
{\Large\bf 
On cutoff effects in lattice QCD from short to long distances
\\[0.5ex] 
}
\end{center}
\vskip 0.35cm
\vbox{
\centerline{
\epsfxsize=2.8 true cm
\epsfbox{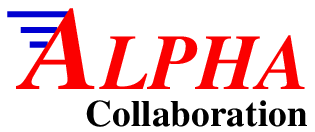}}
}
\vskip 0.1cm
\begin{center}
{
Michele Della Morte$^{\scriptscriptstyle a}$,
Rainer Sommer$^{\scriptscriptstyle b}$ and
Shinji Takeda$^{\scriptscriptstyle c}$
}
\vskip 0.5cm
{
\vskip 2.0ex
$^{\scriptstyle a}$
CERN,
Physics Department, TH Division, CH-1211 Geneva 23, Switzerland
\vskip 2.0ex
$^{\scriptstyle b}$
DESY,
Platanenallee 6, 15738 Zeuthen,  Germany
\vskip 2.0ex
$^{\scriptstyle c}$
Institut f\"ur Physik, Humboldt Universit\"at, Newtonstr. 15, 12489 
Berlin, Germany
\vskip 2.0ex
}
\vskip 0.775cm
{\bf Abstract}
\vskip 0.1ex
\end{center}
We discuss kinematical enhancements of cutoff effects
at short and intermediate distances. Starting from a
pedagogical example with periodic boundary conditions,
we switch to the case of the \SF, where the theoretical
analysis is checked by precise numerical data with
$\nf=2$ dynamical $\Oa$-improved Wilson quarks. Finally
we present an improved determination of the 
renormalization of the axial current in that theory.


\vskip 0.65cm

\begin{center}
July 2008 
\end{center}

\eject
\vfill
\eject

\end{titlepage}

\section{Introduction \label{s:intro}}

Dynamical fermion simulations with Wilson-type fermions are 
now possible at small quark masses, large 
volumes~\cite{algo:GHMC,algo:L2,algo:stability,algo:L3,cern:II,etmc:2007a,Kuramashi:2007gs,bmw:scaling} 
and small lattice spacings. With statistical precision reaching the (sub-)percent 
level, an important 
uncertainty which
remains to be carefully controlled  is due to the finite lattice spacing $a$. 
In the non-perturbatively $\Oa$-improved theory, this issue has been investigated
in some detail. On the one hand, significant $a$-effects
have been found at lattice spacings of around $a=0.1\,\fm$ \cite{lat03:rainer,impr:za_nf2}. 
On the other hand, both in the  high precision computations
of the scale dependence of coupling, quark masses and other composite
operators~\cite{alpha:nf2,mbar:nf2,zastat:nf2}  (see ref.~\cite{nara:rainer} for a review) 
and in a recent scaling test 
\cite{scaling:nf2}, such effects
were not visible.

In this paper we analyse this apparent contradiction and find that the difference is
of a simple kinematical origin, teaching us a more general lesson. In order to keep 
$a$-effects small, it is important to 
ensure that the general 
conditions necessary for the application of the Symanzik expansion of lattice 
observables in powers of $a$ are well fulfilled. We discuss the general
issue in  \sect{s:general}
and then illustrate it in  \sect{s:fp} on a particular example where precise non-perturbative
results are available: 
a \SF correlation function.
The lessons we learn there allow us to perform an improved computation of
the renormalization of the axial current in \sect{s:za}, which we check
further by considering modified renormalization conditions~(\sect{s:checks})
before we conclude.

\section{The Symanzik expansion and kinematics
  \label{s:general}}

Symanzik's effective theory\cite{impr:Sym1,impr:Sym2,impr:pap1} 
is the fundamental tool for analysing the approach of 
renormalized lattice observables, $O_\mrm{lat}$ to their continuum 
limit $O_\mrm{cont}$. 
As an example consider the asymptotic expansion of a (space-momentum zero)
correlation
function (e.g. \eq{e:fpp})
\def\corr#1{{C}_{\rm #1}}
\bes 
    \corr{lat}(x_0, a) &\simas{a\to0}&
     \corr{cont}(x_0) 
    + a\, \corr{1}(x_0) 
    \nonumber 
    + a^2\,\corr{2}(x_0) \, 
    + \ldots \,.
\ees 
The functions $\corr{}$ depend in addition to the 
kinematical argument $x_0$ on the 
intrinsic scale $\Lambda \equiv \Lambda_\mrm{QCD} = \rmO(300\,\MeV)$
of the theory as well as the quark masses $M_i$. 
In the above formula the powers of the lattice spacing,
are in principle modified
by a further $a$-dependence in 
$C_i(x_0) \to  C_i(x_0, a)$.
It is a consequence of the dependence of the couplings
in the Symanzik effective Lagrangian on $a$ as well as the
dependence of the coefficients of the effective fields
on $a$. However, since QCD is asymptotically free, these 
coefficients depend only logarithmically on the lattice spacing
when it is small. Such logarithmic terms are
of  minor importance for our discussion; they are dropped here. 

\subsection{Point-to-point correlators at short distances\label{s:pp} }

Let us assume a large volume and a ``point-to-point'' correlator,
i.e. $\corr{lat}(x_0, a) = a^3 \sum_{\vecx}\langle \op{1}(0) \op{2}(x)\rangle$
with local operators $\op{1,2}$. 
We focus on the  short distance regime, $x_0 \ll 1/\Lambda\,,
x_0 \ll 1/M_i$ and
an $\Oa$-improved theory where $ \corr{1} \equiv 0$.  Then $x_0$ 
provides the only dimensionful parameter. 
On purely dimensional ground the leading correction then becomes
\bes
    a^2 \, \corr{2}(x_0) 
    &\simas{a\to0}& \mrm{const.}\,\times\, (a/x_0)^{2}\, \corr{cont}(x_0)\,.
    \label{e:corr2}
\ees
It is enhanced at small $x_0$. In other words what we have used here 
is that up to logarithmic corrections the short distance correlation 
functions have a unique power-law behaviour in $x_0$. 

A particularly
simple example is 
\bes
  \label{e:fpp}
  \corr{lat}^\mrm{PP}(x_0,a) = 
  - a^3 \sum_{\vecx}\langle P^a(0)\, P^a(x)\rangle \,, 
  \quad \mbox{with} \quad
  P^a(x)=\psibar(x) \gamma_5 \frac12 \tau^a \psi(x)\,,
\ees
which one may, in principle, consider for 
renormalizing the pseudo--scalar density.  One expects 
that the short distance behaviour
$\corr{cont}^\mrm{PP} \sim \mrm{const.}\,\times x_0^{-3}$
is difficult to 
reproduce accurately on a lattice. In fact, {\em one} term in the 
Symanzik expansion originates from an $\rmO(a^2)$ correction
to the field
\bes
  P^a_\mrm{eff}(x) =  P^a(x) + a^2 c_\mrm{2}\, \partial^*_\mu \partial_\mu P^a(x)+ \ldots \,
\ees
which contributes 
\bes
    \corr{2}^\mrm{PP}(x_0)\, a^2 
    &\simas{a\to0}& 24\, c_\mrm{2} { a^2 \over x_0^2}\, \corr{cont}^\mrm{PP}(x_0) 
    + \ldots 
    \label{e:corr22}\,
\ees
to  \eq{e:corr2}. 
An order of magnitude enhancement is due to the second derivative of the steep function
$\corr{cont}$. Even if $c_\mrm{2}$ may be arising only at 
1-loop of perturbation theory\footnote{this means  $c_2 \sim  \mrm{const.}\,/\log(a \Lambda)$}, 
the considered case suffices to illustrate our main point: cutoff effects may have a 
significant kinematical
enhancement. Particular examples are correlation functions
with strong short distance singularities.

Of course, this is the reason why the connection between 
the perturbative short distance regime of QCD and the 
non-perturbative long distance one
is carried out recursively in the strategy of our collaboration 
\cite{alpha:sigma,alpha:su2,nara:rainer}. It is then possible to
have $a/x_0 \ll 1$ {\em and} $x_0\Lambda \ll 1$.  But furthermore, by making use of 
\SF boundary conditions\cite{SF:stefan1}, one may easily construct correlators 
with a weak time-dependence
\cite{mbar:pap1}. Let us discuss this relevant issue in some more detail.

\subsection{\SF correlators}

We now assume a finite volume with
Dirichlet boundary conditions at $x_0=0$ and $x_0=T$
as explained in refs.~\cite{SF:LNWW,SF:stefan1,impr:pap1}. 
These allow for the definition of gauge invariant boundary fields,
for instance 
\bes
  {\mathcal{O}}^a = {a^6 \over L^3}  \sum_{\vecx} \sum_{\vecy} 
  \zetabar(\vecx)\, \gamma_5\,\frac{1}{2}\tau^a\,  \zeta(\vecy)
  \label{e:boundop}
\ees
constructed from the boundary quark (and anti-quark) fields
$\zeta$ ($\zetabar$) at $x_0=0$. In space we use periodic
boundary conditions with a phase $\theta$,
\bes
  \psi(x+\hat{k}L) = \rme^{i \theta}\, \psi(x)\,,\quad
  \psibar(x+\hat{k}L) = \rme^{-i \theta}\, \psibar(x)\,. 
\ees
Because of the zero momentum projection of the 
boundary fields, the correlator
\bes
  f_\mrm{lat}^\mrm{P}(x_0, a) =  -\,{L^3 \over 6} \langle {\cal O}^a\, P^a(x)\rangle
\ees
has a smooth behaviour at short distances
(and small $L$) of the form
\bes
  \label{e:fppert}
  f_\mrm{cont}^\mrm{P}(x_0) =  3 \rme^{-2 \sqrt{3} \theta\,x_0/L} \,
                   \left\{ 1 + \rmO(\gbar^2(L) \right\}\,,
\ees
in QCD with 3 colors and for massless quarks.
There is no kinematical enhancement of $a$-effects. 
Together
with the smooth background field introduced for the definition of the 
running coupling~\cite{SF:LNWW,alpha:su3}, this explains the very 
small lattice spacing effects in the running coupling and
running operators mentioned before. 

We may also discuss the behaviour at large distances.
There a saturation by few intermediate states in the
spectral decomposition (see ref.~\cite{mbar:pap2} for a discussion
of $ w,E$)
\bes
 \label{e:spect}
 f_\mrm{lat}^\mrm{P}(x_0) = \mrm{const.}\,\sum_{n,m}
              w_{nm} \rme^{-x_0 E_n^{\pi} 
                     - (T-x_0)\,E_m^\mrm{vac} }  
\ees 
will give an accurate description of the correlation
function. The energies $E_n^\pi$ are the finite volume
eigenvalues of the QCD Hamiltonian in the pion sector and
$E_m^\mrm{vac}$ are the energies of states with vacuum quantum
numbers. Their dependence on $L$ and the lattice spacing is 
suppressed. If the kinematics, given by $\theta,x_0,L$ is such that
effectively a few states with energies up to $a E \approx 1$ contribute, 
the $x_0$-dependence may again be strong and
cutoff effects may be enhanced. However, 
the enhancement will not be as large as in \sect{s:pp}
since there is no singular dependence at small $x_0$.
In \sect{s:fp} we will 
see quantitatively how the behaviour changes as $L$ and $x_0$ are increased
starting from $L, x_0 \ll 1/\Lambda$.

We remark that the foregoing discussion is of course not in 
contradiction to the perturbative behaviour. In the perturbative
region, $x_0$ is small and many states contribute significantly.
But asymptotic freedom implies that their coefficients $w_{nm}$ are fine-tuned 
such as to produce the smooth
behaviour of \eq{e:fppert}. This is often called quark-hadron duality.

\subsection{Energies and matrix elements} 

The most common application of Symanzik's effective theory is to 
energies and matrix elements, e.g. extracted from \eq{e:spect} at
large $x_0$. The expansion of such observables,
\bes
    O_\mrm{lat} \sim O_\mrm{cont} + a\, s_1\, + \,a^2\, s_2 + \ldots\,,
\ees 
will be valid and accurate when the relevant energies, momenta and masses are
small compared to $a^{-1}$. For the $\Oa$-improved theory, a first scaling test 
\cite{scaling:nf2} showed that indeed the corrections to the continuum
appear to be reasonably small when $a\leq 0.1 \, \fm$.

\section{The Schr\"odinger functional correlator $f_\mrm{lat}^\mrm{P}$ 
\label{s:fp}}

\vspace{0.4cm}
\begin{figure}[htb]
\begin{center}
\epsfig{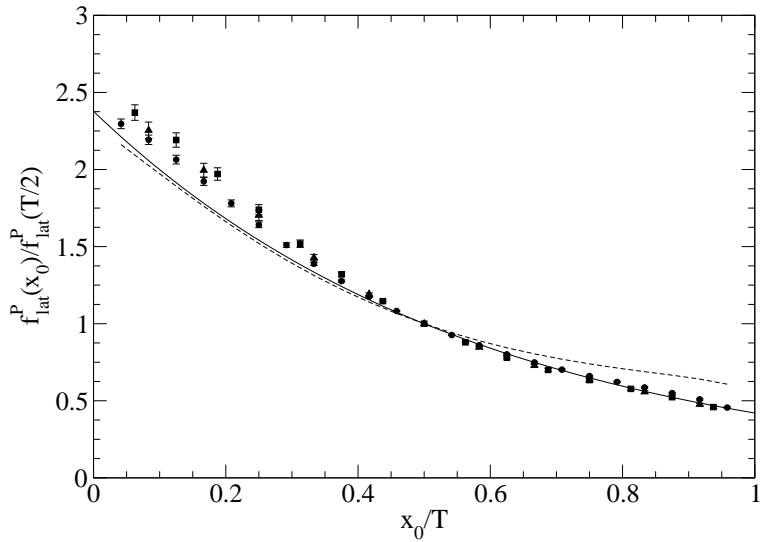}
\end{center}
\caption{\footnotesize{\sl The correlator  $f_\mrm{lat}^\mrm{P}$
       in the perturbative region, $L \leq 0.5\,\fm$.
       The continuous curve shows the tree-level behaviour,
       while data points are for $\gbar^2(L/2)=3.33$ (or $\gbar^2(L)\approx5.5$)
       with filled circles for $L/a=24$, squares for $L/a=16$ and triangle 
       for $L/a=12$.
       A dashed line shows the position of the data with $\gbar^2(L/2)=1.50$, 
       $L/a=24$.
} }
\label{f:fppert}
\end{figure}

\begin{figure}[htb]
\begin{center}
\epsfig{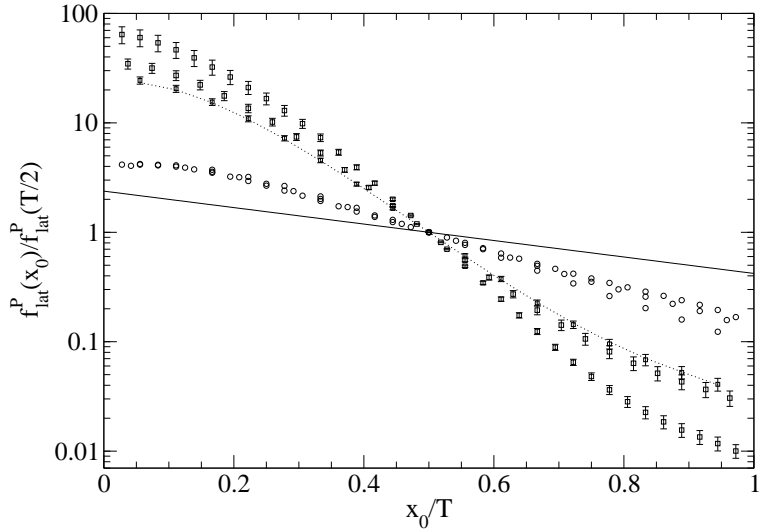}
\end{center}
\caption{\footnotesize{\sl The correlator  $f_\mrm{lat}^\mrm{P}$
         for $L \approx 0.8\,\fm $, 
         $T\approx 1.8\,\fm$, 
         $\theta=0.5$ is represented by squares. $L/a=8,12,16$ data 
         increase at small $x_0$.
         A dotted line indicates the location of  $\theta=0, L/a=8$.
         Circles show the behaviour at $L \approx 1.2\,\fm $, 
         $T\approx 1.8\,\fm$, 
         $\theta=0$. Their errors are of the order of
         the symbol size. 
} }
\label{f:fpintermed}
\end{figure}

We now proceed to discuss
numerical results for $\fp$ in different regions of $L$ (or $\gbar^2(L)$). 
Let us start from the short distance, weak coupling, region.
In the non-perturbative computation of the scale-dependence
of composite operators, such as $P^a$, our collaboration 
chose \SF boundary conditions with $\theta=0.5$, $T=L$ and a vanishing
background field \cite{mbar:nf2}. From these simulations we 
have precise data for $f_\mrm{lat}^\mrm{P}$ in a range of the \SF coupling 
$0.98 \leq \gbar^2(L) \leq 5.5$  \cite{alpha:nf2,mbar:nf2}. In order
to cancel the renormalization factor $\zp$, we normalize by 
 $f_\mrm{lat}^\mrm{P}$ at the midpoint, $x_0=T/2=L/2$. 

At $\gbar^2=0$,
\eq{e:fppert} yields the slowly falling exponential 
plotted in \Fig{f:fppert} as a continuous curve. 
The non-perturbative results at $\gbar^2(L) \approx 5.5$
are next to this tree-level curve, even though in general perturbation theory 
is not very accurate at this coupling. For example the renormalization
factor $\zp$ is known to be a factor 2 away from one. We note in passing
that the {\em very} close agreement 
with tree-level at $x_0/T \geq 1/2$ is somewhat accidental, as e.g. 
at a weak coupling of $\gbar^2(L/2)=1.50$ the curve is some 15 \% away (dashed line). 
All in all, the smooth behaviour anticipated in our general discussion, is found
to a remarkable degree in the whole range of $L\,\ltsim\, 0.5\,\fm$. 
Cutoff effects are very small.
Even at the shortest distances, $x_0/a$, the $a^2$ effects 
are only a few percent at $\gbar^2(L)=5.5$. 
At weaker coupling, $\gbar^2(L/2)=1.50$, the $a^2$ effects are invisible within our 
precision of $\approx 2 \%$ 
(the data are not plotted in order not to clutter the graph). As explained
in the previous section,
this behaviour
of the \SF correlators goes hand in hand with
small cutoff effects in the step scaling functions which describe for example 
the running of coupling and quark mass~\cite{mbar:nf2}. 

In physical units we have thus far investigated the region 
$L=T\,\ltsim\, 0.5\,\fm$. For $L$ just somewhat larger,
$L=0.8\,\fm$, $T=9/4\,L$, and $\theta=0$, we have previously observed 
significant $a^2$
effects in our computation of $\za$ \cite{impr:za_nf2}. They were
prominent
in the statistically significant disconnected contributions to
$\za$, which can be shown to vanish in the continuum limit. As a consequence
$\za$ determined with or without the disconnected diagrams differed
by almost 15\% at $a\approx0.1\,\fm$. Although the disconnected diagrams
vanish quickly as $a$ is reduced, an unpleasantly large ambiguity remained
at typical values of $a$. For details we refer to the quoted reference. 

We now turn to that same kinematics and subsequently to larger $L$. 
But first we point out that in this region the 
dependence of $f_\mrm{lat}^\mrm{P}(x_0)/f_\mrm{lat}^\mrm{P}(T/2)$
on $\theta$ is insignificant compared to the 
effects we will discuss. An explicit example is provided 
by comparing the
dotted line of \Fig{f:fpintermed} with the $L/a=8$ data points. 
For our numerical demonstration
we will thus freely use data at available values of $\theta$.  

We start with correlation functions $f_\mrm{lat}^\mrm{P}$ 
at the same parameters as in \cite{impr:za_nf2}, 
except for that we remain with $\theta=0.5$. 
The behaviour of $f_\mrm{lat}^\mrm{P}$, plotted 
in \Fig{f:fpintermed}, differs drastically from \Fig{f:fppert} -- see the 
tree-level curve as a reference. The non-perturbative
correlators drop steeply 
(note that we use a logarithmic scale) and follow the characteristics 
of the described intermediate regime, where neither the smooth perturbative
behaviour is realized, nor a single intermediate state is 
dominating (the latter would be seen as a straight line in the figure).
For the coarsest lattice the logarithmic slope (effective mass)
at $x_0=T/2$ is as high as $0.8$ in lattice units. In this situation,
we find indeed very significant lattice spacing effects. 
It is then not surprising that there are also large  
$a^2$ effects in the form of the mentioned disconnected diagrams.

For an improved determination of $\za$, we would like to choose
a kinematical region, where (1) the $a^2$ effects are better
suppressed and (2) the \SF simulations can be done close to or in the 
chiral limit. The simulations can reach the  chiral limit when the 
infrared cutoff
on the spectrum of the Dirac operator is sufficiently large.
Since in the \SF the infrared cutoff is 
dominantly controlled by $1/T$, we do not want to increase $T$ 
significantly compared to the previous $T\approx1.8\,\fm$.
On the other hand, the energies
$E_n^\pi$ are expected to be decreasing with $L$; in fact 
in the small $L$ limit they scale as $L^{-1}$. This leads us to
consider a somewhat larger $L$, 
namely $L\approx 1.2\,\fm$. 
Also in such a situation, namely with
 $T = 3/2\,L$, we
have simulation results \cite{impr:babp_nf2}, generated for the 
determination of improvement
coefficients $\bm, \,\ba-\bp$ as well as the renormalization of the
bare mass $\mq$ following ref.~\cite{impr:babp}. The circles in
\Fig{f:fpintermed} 
show $f_\mrm{lat}^\mrm{P}$ in this kinematical
situation\footnote{         
         At $L/a=12$ the massless point is not reached, but
         the behaviour at the smallest mass, shown here, 
         does not change significantly when the mass is increased.
         See \sect{s:za} for details.  
}. The function has a much slower decay and indeed only moderate
lattice artifacts. The energies of the states dominating the 
correlator appear to be significantly lower -- at least the
logarithmic slope never exceeds a value of $0.3$ at the coarsest 
lattice spacing. 

This kinematics is a good starting point for
a redetermination of the renormalization factor $\za$ with reduced
intrinsic $a^2$ ambiguities.

\section{New determination of $Z_{\rm A}$\label{s:za}}
The renormalization condition of the axial current is obtained by considering
an axial Ward identity
in exactly the same way as it was done in ref.~\cite{impr:za_nf2}. One starts
from a \SF correlation function of the axial current with two pseudo-scalar 
boundary operators,
\bes
{\mathcal{O}}^a(\omega)&=& {a^6 \over L^3}
\sum_{{\bf x,y}}\zetabar({\bf x})\gamma_5 \frac{1}{2}\tau^a\omega({\bf x}-
{\bf y}) \zeta({\bf y}) 
\\
{\mathcal{O}'}^a(\omega)&=& {a^6 \over L^3}
\sum_{{\bf x,y}}\zetabarprime({\bf x})\gamma_5 \frac{1}{2}\tau^a\omega({\bf x}-
{\bf y}) \zeta'({\bf y})
\ees
and performs an axial rotation of the variables
in a region around the axial current\cite{impr:pap4}. 
Using isospin symmetry and PCAC one obtains for vanishing quark mass the 
Ward identity
\begin{eqnarray}
\za^2 f_\mrm{AA}^\mrm{I}(x_0,y_0,\omega)=f_1(\omega)+\rmO(a^2)\,,\label{normcon1}
\quad x_0>y_0\,,
\end{eqnarray}
in terms of the correlation function
\begin{eqnarray}
f_\mrm{AA}^\mrm{I}(x_0,y_0,\omega)&=&-\frac{a^6}{6}\sum_{\vecx,\vecy}
\epsilon^{abc}\epsilon^{cde}
\Big\langle \mathcal{O'}^d(\omega)\, (\aimpr)_0^a(x)\,(\aimpr)_0^b(y)\,\mathcal{O}^e(\omega)\Big\rangle\,, 
\label{fAAI}
\\
f_1(\omega)&=&-\frac1{3}\langle\mathcal{O'}^a(\omega)\,\mathcal{O}^a(\omega)\rangle\label{f1}\,.
\end{eqnarray}
The superscript $\mrm{I}$ reminds us that the improved axial current
\begin{equation}
\label{imprcurrent}
(\aimpr)_\mu^a=A_\mu^a+a \ca\drvsym\mu{} P^a\,, \quad A_\mu^a(x)=\psibar(x)\gamma_\mu\gamma_5\frac{1}{2}\tau^a\psi(x)\,,
\end{equation}
is to be inserted. 
We refer to section 2 in ref.~\cite{impr:za_nf2} for a derivation,
the exact lattice implementation as well as a generalization to 
finite mass, which we use.

Compared to that work we only change the kinematics. 
First of all the computation is now performed on a lattice of size 
$L\simeq 1.2$~fm instead of $0.8$ 
fm, as motivated in the previous sections.
Secondly, together with the sources ${\mathcal{O}}^a$,  
\eq{e:boundop}, and ${\mathcal{O}}'^a$, which correspond to $\omega_0(\vecx)=1$,
we consider the following basis of 
wave-functions
\bea
\omega_i(\vecx) &=& N_i^{-1} \sum_{{\bf n}\in{\bf Z}^3}
\overline{\omega}_i(|\vecx-{\bf n}L|)\,,\; i=1,2,3\,,
\\
\overline{\omega}_1(r) &=& \rme^{-r/a_0}\,,\quad \nonumber 
\overline{\omega}_2(r) =  r\,\rme^{-r/a_0} \,,\quad
\overline{\omega}_3(r) = \rme^{-r/(2a_0)}\,\;,
\label{eq:wf}
\eea
in order to construct the external operators. 
Here, we keep the physical length scale
$a_0$ fixed in units of $L$ by  choosing $a_0\!=\!L/6$ and
the (dimensionful) coefficients $N_i$ are set to normalize the wave function via
$a^3\sum_{\vecx} \omega_i^2({\bf x})=L^3$. 
The same set of interpolating fields, for the same $L$,
has been used in ref.~\cite{impr:ca_nf2} to determine the improvement
coefficient $c_{\rm A}$ by requiring the quark mass derived from the PCAC  Ward identity 
to stay the same
as the external states are changed. We therefore choose the 
already determined~\cite{impr:ca_nf2} optimal 
wave-function
\bes
  \omega_{\pi^{(0)}} = \sum_{i=1}^3\eta_i^{(0)}\omega_i\,,\quad \eta^{(0)}=(0.5172,\, 0.6023,\,0.6081)\,.
\ees
It suppresses the 
contribution of the first excited state in the pseudoscalar channel to the correlation functions under
consideration (see section 2 in ref.~\cite{impr:ca_nf2}).

The final result will turn out to differ significantly from the determination in ref.~\cite{impr:za_nf2} 
at the two largest couplings only. We therefore did not recompute $Z_{\rm A}$ for small couplings and
rather use the old estimates.
\subsection{Results}
We have two dynamical flavors of non-perturbatively improved Wilson quarks~\cite{impr:csw_nf2} and the 
plaquette gauge action. The improvement coefficients $\csw,\ca$ were set
to their non-perturbative values\cite{impr:csw_nf2,impr:ca_nf2}. We chose $T=3/2L$ with periodic boundary 
conditions ($\theta=0$) in space and vanishing background field. 
We used the HMC algorithm with two pseudo-fermion
fields as proposed in refs.~\cite{algo:GHMC,Hasenbusch:2002ai}. The particular implementation 
has been discussed 
and tested in refs.~\cite{algo:GHMCalpha} and~\cite{algo:trajlength}.
Following the last reference we chose a trajectory length of $\tau=2$ except
for at $\beta=5.2$ where we set $\tau=1$.

The normalization factor $Z_{\rm A}$ is given
by~\cite{impr:za_nf2}
\begin{equation}
Z_{\rm A}(g_0^2)=\lim_{m \to 0} \sqrt{f_1(\omega)}\left[f_{\rm AA}^{\rm I}(2T/3,T/3,\omega)
-2m\tilde{f}_{\rm PA}^{\rm I}(2T/3,T/3,\omega) \right]^{-1/2}\;. 
\end{equation}
The definition of $\tilde{f}_{\rm PA}^{\rm I}(2T/3,T/3,\omega)$ and the PCAC quark mass $m$ 
can be found in ref.~\cite{impr:za_nf2}. 
After performing the Wick contractions one realizes that disconnected quark diagrams, where no propagator
connects the two boundaries, contribute to $f^{\rm I}_{\rm XY}(x_0,y_0,\omega)$.
These can be shown to vanish in the continuum massless limit as a consequence of the conservation of
the axial current~\cite{impr:za_nf2}. 
In an improved theory they therefore amount to O$(a^2)$ effects on $Z_{\rm A}$.
By dropping them one obtains an alternative definition of $Z_{\rm A}$, denoted $Z_{\rm A}^{\rm con}$. 
With the kinematics of ref.~\cite{impr:za_nf2} the difference between $Z_{\rm A}$ and $Z_{\rm A}^{\rm con}$
for $\beta=6/g_0^2<5.5$ was found to be rather large, though consistent with O$(a^2$) scaling.
As it will become clear in the  following this effect is very small in the computation 
presented here, which therefore significantly improves on the result in ref.~\cite{impr:za_nf2}.

In order to ensure a smooth dependence of $Z_{\rm A}$ on the bare coupling $g_0^2$ and
the correct scaling of discretization errors proportional to $a^2$, we impose our normalization
condition on a line of constant physics. This requires keeping all length scales fixed as $g_0^2$ is 
varied.
The lattice size $L$ is set to approximately $1.8L^*$ with $L^*$ given by the condition 
$\gbar^2(L^*)= 5.5$, where $\gbar^2(L)$ is the Schr\"odinger functional coupling defined in 
refs.~\cite{alpha:su3,alpha:nf2}. The relation between $\frac{L^*}{a}$ and $g_0$ 
could be taken from 
ref.~\cite{lat07:rainer}.

\begin{figure}[h!]
\begin{center}
\epsfig{file=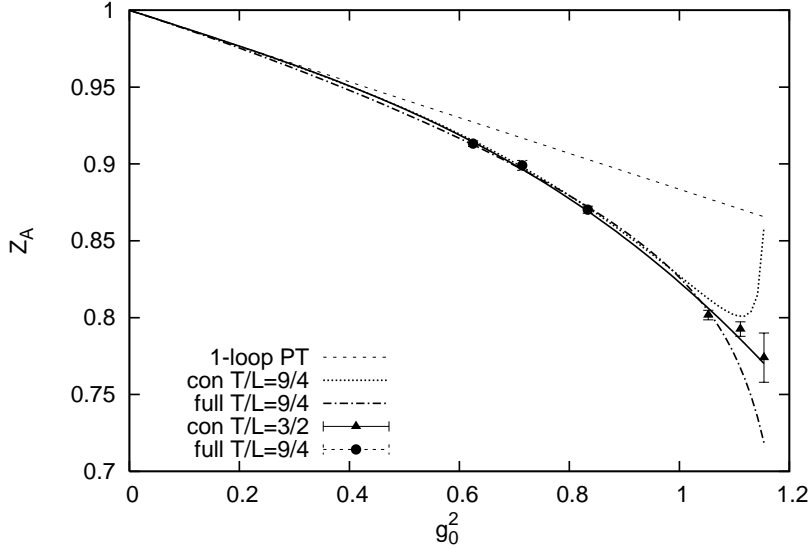,width=11cm}
\end{center}
\caption{\footnotesize{\sl Comparison of the new results for $Z_{\rm
 A}^{\rm con}$ with the old ones for $Z_{\rm A}$ and $Z_{\rm A}^{\rm con}$ in
ref.~\cite{impr:za_nf2}.
The solid curve represents the interpolation formula in eq.~(\ref{ZAform}).
}}
\label{f:ZAplot}
\end{figure}

The bare parameters of 
our simulations and the results for $Z_{\rm A}$ are collected in Table~\ref{ZAtable}.
{\tiny
\begin{table}[htb]
\centering
\begin{tabular}{cccc|cc|cc}
\hline\\[-2.5ex]
   &   &     &    & \multicolumn{2}{c|}{$\omega=\omega_0$} & \multicolumn{2}{c}{$\omega=\omega_{\pi^{(0)}}$} \\[1ex]
$L/a$ & $\beta$ & $\kappa$  & $am$  & $Z_{\rm A}$ & $Z_{\rm A}^{\rm con}$ & $Z_{\rm A}$ & 
$Z_{\rm A}^{\rm con}$  \\[1ex]  
\hline 
    12 & 5.2   & 0.1355    &  0.02121(36) &0.788(15)  &0.784(4) & 0.784(16)  & 0.7874(35)  \\
    12 & 5.2   & 0.1357    &  0.01434(48) &0.775(10)  &0.769(4) & 0.777(11)  & 0.7703(36)  \\
    12 & 5.2   & 0.1358    &  0.00907(39) &0.776(~8)  &0.777(4) & 0.776(9) & 0.7788(37)  \\
    12 & 5.2   &           &  $\to 0$     &0.766(18)  &0.773(15)  & 0.769(20)  & 0.774(16)   \\
    16 & 5.4   & 0.136645  &  0.00062(26) &           &           & 0.779(5) & 0.793(5)  \\
    24 & 5.7   & 0.136704  &  0.00072(14) &           &           & 0.808(5) & 0.802(3)  \\[1ex]
\hline
\end{tabular}
\caption{\footnotesize{\sl Bare parameters and simulation results (statistical errors only). 
At $\beta=5.2$ the value of the renormalization constant extrapolated to the chiral limit is reported in the 
fourth line including the associated systematic uncertainty (see text for details). The number of decorrelated
measurements used varies between $1200$ (at $\beta=5.2$) and $200$ (at $\beta=5.7$).
}}
\label{ZAtable}
\end{table}
}
Due to algorithmic instabilities caused by the appearance of very small, unphysical, eigenvalues in the spectrum 
of the Wilson-Dirac (SF) operator~\cite{algo:cutoff,algo:stability}, 
at the coarsest lattice spacing we could simulate down to bare quark masses of about $a\,m\approx0.01$ only. 
\begin{figure}[h!]
\begin{center}
\epsfig{file=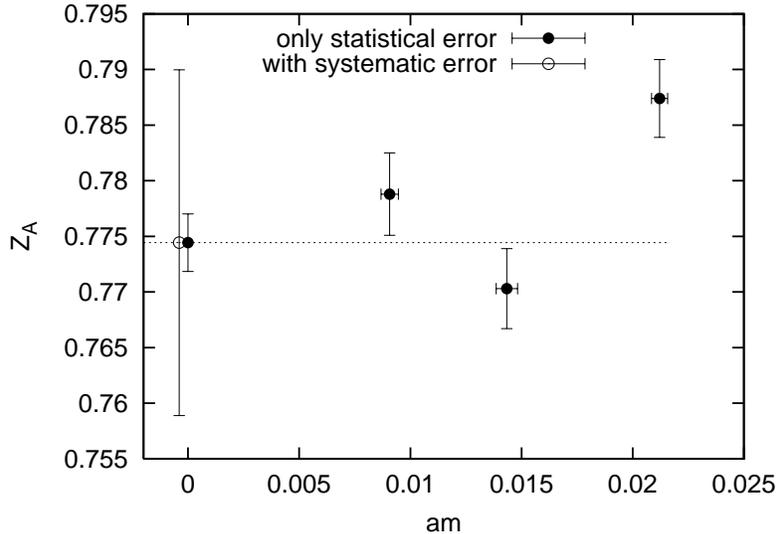,width=10.5cm}
\end{center}
\caption{\footnotesize{\sl Estimate of
$Z_{\rm A}^{\rm con}$ at the chiral point for $\beta=5.2$.
The error-bar of the filled circle at $am=0$ shows
the statistical error whereas the error on the open circle 
includes our estimate of the $\rmO(am)$ systematic uncertainty.
}}
\label{f:ZAextrap}
\end{figure}
Our estimates for the  massless limits at $\beta=5.2$ 
are just weighted averages of the numbers at the two lightest
quark masses. The errors in Table~\ref{ZAtable} include
a systematic uncertainty for possible $\rmO(am)$ contaminations given by the difference between the 
determination
at the heaviest mass and the described estimate.
This uncertainty is added linearly to the statistical error. A similar $\rmO(am)$ uncertainty would be
obtained by comparing to a linear fit in $am$ with all three points.
Figure~\ref{f:ZAextrap}
illustrates the procedure for $Z_{\rm A}^{\rm con}$ and $\omega=\omega_{\pi^{(0)}}$.
It is clear from the plot that the dependence of $Z_{\rm A}$ on the quark mass is rather mild as expected for 
the ``massive''definition of the renormalization constant~\cite{impr:za_nf2} used here.
Similar remarks obviously apply to the case $\omega=\omega_0$, as \tab{ZAtable} shows that
the two wave-functions 
give consistent results for all values of $\kappa$.
At the other values of the bare coupling we simulated at very small quark masses.
Given also the flat dependence 
observed at $\beta=5.2$ we did not need to estimate an effect of order $am$.
As anticipated, already at $\beta=5.7$ the present result nicely agrees with the numbers in 
ref.~\cite{impr:za_nf2}. Conversely, at $\beta=5.2$ the new determination of $Z_{\rm A}$ is about 
8\% larger than the old one. These comparisons are summarized in Figure~\ref{f:ZAplot}.

In view of the above discussion we smoothly interpolate the data for $Z_{\rm A}^{\rm con}$ and $\omega=\omega_{\pi^{(0)}}$in the region 
$5.2 \leq \beta \leq 5.7$ and those in ref.~\cite{impr:za_nf2} for $7.2 \leq \beta \leq 9.6$
(filled circles in Figure~\ref{f:ZAplot}) by the formula
\be
Z_{\rm A}(g_0^2)=1-0.116g_0^2 +0.011g_0^4-0.072g_0^6 \;,
\label{ZAform}
\ee
where the coefficient of the term linear in $g_0^2$ is fixed by 1-loop
perturbation theory~\cite{pert:gabrielli} and the last two coefficients are the result of  a fit.
We ascribe an absolute error to $Z_{\rm A}$, which decreases from 0.016 at $\beta=5.2$ over
0.005 at $\beta=5.4$ to 0.003 for
$\beta\geq 5.7$. We note that our systematic $\rmO(am)$ error at $\beta=5.2$ is rather conservative,
see \fig{f:ZAextrap}.


\subsection{Other renormalization conditions of the 
  axial current \label{s:checks}}
In order to get a further impression about residual cutoff effects, 
we also studied an alternative definition of  $Z_{\rm A}$.
It is obtained in the same framework by 
replacing a light quark with a static one in the external operators ${\mathcal{O}}$ and
${\mathcal{O}}'$. By remaining with the flat wave-function and by denoting the static quark field
on the boundary $x_0=0$  by $\zeta_{\rm h}$ we write
\bea
{\mathcal{O}}^{\rm hl}_i={a^6\over L^3} \sum_{\bf x,y}\zetabar_{\rm h}({\bf x}) \gamma_5
\zeta_i({\bf y}) \,,&& 
{\mathcal{O}}'^{\rm hl}_i={a^6\over L^3} \sum_{\bf x,y}\zetabar'_i({\bf x}) \gamma_5
\zeta'_{\rm h}({\bf y}) \;, 
\label{e:boundopstat}
\eea
where $i=1,2$ is a flavour index. The flavour contractions in $f_{\rm XY}^\mrm{I}$ are 
changed correspondingly. The correlator $f_{\rm AA}^\mrm{hl,I}$ for example is written
\bea
f_{\rm AA}^\mrm{hl,I}(x_0,y_0)&=&-{{ia^6}\over{6}}\sum_{\bf x,y} \epsilon^{abc} \langle 
{\mathcal{O}}'^{\rm hl}_i (\aimpr)_0^a(x)\,(\aimpr)_0^b(y) \frac{1}{2}(\tau^c)_{ij} {\mathcal{O}}^{\rm hl}_j \rangle
\eea
and the massless normalization condition becomes
\be
Z_{\rm A}^2\,f_{\rm AA}^\mrm{hl,I}(2T/3,T/3) =f_1^{\rm hl} +{\rm O}(a^2) \;,
\ee
with 
\be
f_1^{\rm hl}=-{{1}\over{8}}\,\langle{\mathcal{O}}'^{\rm hl}_i {\mathcal{O}}^{\rm hl}_i\rangle 
\;.
\ee
As the fields $X$ and $Y$ in the correlator
do not contain static fields,
it is clear that disconnected diagrams cannot appear and the 
static quark propagates from one boundary to the other.
Static quarks are discretized through the HYP1 and HYP2 static-quark actions~\cite{Della Morte:2005yc}, which 
have a relatively good signal to noise ratio in static-light correlation functions 
at large time separations.

{\tiny
\begin{table}[htb]
\centering
\begin{tabular}{cccccccc}
\hline\\[-2.5ex]
$L/a$    &  $T/a$ &   $\beta$    & $\kappa$   &  $am$ &   $\theta$  &  $Z_{\rm A}\,$[HYP1] & $Z_{\rm A}\,$[HYP2] \\[1ex]
\hline
12       &   18   &     5.2      &  0.1355    &0.02121(36)&   0     &          0.7904(66)  & 0.7882(55) \\
12       &   18   &     5.2      &  0.1357    &0.01434(48)&   0     &          0.7729(79)  & 0.7734(73) \\
12       &   18   &     5.2      &  0.1358    &0.00907(39)&   0     &          0.7731(75)  & 0.7736(70) \\
8        &   18   &     5.2      &  0.1357    &0.00662(75)&  0.5    &          0.8284(80)  & 0.8285(66) \\
8        &   18   &     5.2      &  0.1358    &0.00325(82)&  0.5    &          0.8252(94)  & 0.8206(82) 
\\[1ex]
\hline
\end{tabular}
\caption{\footnotesize{\sl Simulations parameters and results for the alternative "static" definition 
of $Z_{\rm A}$.}}
\label{t:staticza}
\end{table}
}
The check was performed at the largest lattice spacing where ambiguities 
are expected to be most pronounced.
Simulation parameters and results are collected in Table~\ref{t:staticza}.
The results at $L\simeq 1.2$ fm agree with those in the 
previous section, indicating again small overall cutoff effects. 
The values at $L\simeq 0.8$ fm instead again suggest that the determination in  ref.~\cite{impr:za_nf2}
suffers from large $a^2$ effects, as both $Z_{\rm A}$ and $Z_{\rm A}^{\rm con}$ there differ significantly 
(and especially for small quark masses) from the numbers in Table~\ref{t:staticza}.

\subsection{ Renormalization of the vector current\label{s:zv}}
We also recomputed the renormalization constant $\zv$ of the vector current
in the new kinematics. Since it changes by less than 2\% compared to
ref.~\cite{impr:za_nf2} at the largest lattice spacing,
there is no reason to publish a new determination.

\section{Conclusions \label{s:6}}

We have discussed possible kinematical enhancements of 
cutoff effects in lattice gauge theory determinations
of QCD correlation functions. In this respect, the typical
SF correlation functions improve significantly over
those of composite local fields with periodic boundary
conditions (or large volume). The origin of this difference
is simply the difference in mass dimensions of the 
\SF boundary fields compared to the usual local composite 
fields. Since the latter is at least three, time-slice
correlators diverge at least as $x_0^{-3}$ at short distances.
Such a steep behavior is difficult to approximate in a 
discretized theory. 

In the \SF we identified a parametrically much weaker, but still relevant, 
kinematical enhancement of cutoff effects. It appears in
the transition region between approximately perturbative
behaviour and dominantly non-perturbative one. Numerically we find
that it appears e.g. for a $T\times L^3$ geometry with 
$L\approx 0.8\,\fm\,,\; T\approx 1.8\,\fm$. In this kinematical
situation a few hadronic intermediate
states are relevant for the correlators and can produce a 
relatively steep decay of correlation functions even at a small quark mass. 

This means that the transition region from approximately perturbative
to strongly non-perturbative is a relatively difficult one for numerical
simulations. Discretization errors have to be investigated carefully. 
Fortunately we also saw that this region is rather narrow. 
With the step scaling method\cite{alpha:sigma,alpha:su2,mbar:pap1} 
it is typically bridged by one step.

Avoiding the difficult region in the renormalization condition for
the relativistic axial current, we have finally presented a significant
improvement of the previous
determination of $\za$~\cite{impr:za_nf2}. The difference to the old
one is of order $a^2$, but it is up to 10\,\% at the largest lattice
spacing considered. In our new determination we find that 
disconnected contributions, which can be shown to vanish in the 
continuum limit, are very small -- in contrast to the previous
kinematical setup \cite{impr:za_nf2}. We also see a nice agreement with 
a renormalization condition where a static quark is present as a 
spectator in the Ward identity.
The previous determination of $\zv$ is confirmed.

\vspace{0.4cm}
{\bf Acknowledgements.} 
We thank Roland Hoffmann for his collaboration in this project in an
early phase and Ulli Wolff for comments on a first draft of the
paper. 
We acknowledge useful discussions with Martin L\"uscher and 
Ulli Wolff and
thank our colleagues G. de Divitiis, P.~Fritzsch, J.~Heitger, N. Tantalo
and R.~Petronzio for collaborating on the generation of some of the 
\SF gauge configurations. 
We thank NIC  for allocating computer time on the APE
computers at DESY Zeuthen to this project and the APE group for its help. This
work is supported by the  Deutsche Forschungsgemeinschaft 
in the SFB/TR~09  and by 
the European community through 
EU Contract No.~MRTN-CT-2006-035482, ``FLAVIAnet''.


\providecommand{\href}[2]{#2}\begingroup\raggedright\endgroup

\end{document}